\documentclass[twocolumn]{aastex631}

\usepackage{graphicx}	
\usepackage{amsmath}	
\usepackage{amssymb}	

\usepackage{amsmath}
\usepackage{bm}
\usepackage{newtxtext,newtxmath}

\usepackage{graphicx}	
\usepackage{grffile}
\usepackage{amsmath}	
\usepackage{amssymb}	

\usepackage{epsfig,amsmath,natbib}
\usepackage{color,subfigure}
\usepackage{xcolor}

\usepackage{mathrsfs,amssymb,amstext}
\usepackage{ulem}
\usepackage{url}
\usepackage{bm}

\usepackage[T1]{fontenc}

\DeclareRobustCommand{\VAN}[3]{#2}
\let\VANthebibliography\thebibliography
\def\thebibliography{\DeclareRobustCommand{\VAN}[3]{##3}\VANthebibliography}

\def\mean#1{\left< #1 \right>}

\newcommand{\blue}[1]{\textcolor{blue}{#1}}

\begin{document}

\title{A new method of reconstructing Galactic three-dimensional structures using ultralong-wavelength radio observations}

\correspondingauthor{Bin Yue; Yidong Xu; Xuelei Chen}
\email{yuebin@nao.cas.cn; xuyd@nao.cas.cn; xuelei@cosmology.bao.ac.cn}

\author{Yanping Cong}
\affiliation{National Astronomical Observatories, Chinese Academy of Sciences, 20A Datun Road, Chaoyang District, Beijing 100101, China}

\affiliation{School of Astronomy and Space Science, University of Chinese Academy of Sciences, Beijing 100049, China}

\author{Bin Yue}
\affiliation{National Astronomical Observatories, Chinese Academy of Sciences, 20A Datun Road, Chaoyang District, Beijing 100101, China}

\author{Yidong Xu}
\affiliation{National Astronomical Observatories, Chinese Academy of Sciences, 20A Datun Road, Chaoyang District, Beijing 100101, China}

\author{Yuan Shi}
\affiliation{National Astronomical Observatories, Chinese Academy of Sciences, 20A Datun Road, Chaoyang District, Beijing 100101, China}

\affiliation{School of Astronomy and Space Science, University of Chinese Academy of Sciences, Beijing 100049, China}

\author{Xuelei Chen}
\affiliation{National Astronomical Observatories, Chinese Academy of Sciences, 20A Datun Road, Chaoyang District, Beijing 100101, China}

\affiliation{School of Astronomy and Space Science, University of Chinese Academy of Sciences, Beijing 100049, China}

\affiliation{Department of Physics, College of Sciences, Northeastern University, Shenyang, 110819, China}

\affiliation{ Center of High Energy Physics, Peking University, Beijing 100871, China}

\shorttitle{Reconstructing Galactic 3D structures from ultralong-wavelength observations }
\shortauthors{Cong et al.}



\begin{abstract}

The free-free absorption of low frequency radio waves
by thermal electrons in the warm ionized medium of our Galaxy becomes very significant at $\lesssim 10$ MHz (ultralong-wavelength), and the absorption strength depends on the radio frequency. 
Upcoming space experiments such as the Discovering Sky at the Longest wavelength (DSL) and Farside Array for Radio Science Investigations of the Dark ages and Exoplanets (FARSIDE) will produce high-resolution multi-frequency sky maps at the ultralong-wavelength, providing a new window to observe the Universe. In this paper we propose that from these ultralong-wavelength multi-frequency maps, the three-dimensional distribution of the Galactic electrons can be reconstructed. This novel and robust reconstruction of the Galactic electron distribution will be a key science case of those space missions.
Ultralong-wavelength observations will be a powerful tool for studying the astrophysics relevant to the 
 Galactic electron distribution,
for example, the impacts of supernova explosions on electron distribution, and the interaction between interstellar atoms and ionizing photons escaped from the HII regions around massive stars. 
An animation shows the reconstructed results using {\tt NE2001} model as input test. On ArXiv, it is given in the directory: Ancillary files. In the paper the animation is linked to Fig. 5.
\end{abstract}



\section{Introduction}

Understanding our neighborhood and environment has long been a human pursuit,  and the vicinity of the Solar system is of great interest, as it has  direct impacts on the Solar system. Besides the well-observed nearby stars \citep{Gaia2022_asymmetric_disc}
the space is permeated with the interstellar medium (ISM), which serves as the reservoir of the material from which stars formed. Our Sun resides in a low-density cavity named ``Local Bubble'' (LB) 
or ``Local Hot Bubble'' (LHB, \citealt{Cox1987}), 
which is filled with X-ray-emitting hot gas \citep{Snowden1998ApJ}. 
The LB has a typical electron number density of $\sim 5\times10^{-3} $ cm$^{-3}$, which is only 
$\sim20\%-50\%$ of the average value at that radius on the Galactic disk \citep{Ocker2020ApJ}, and it extends to $\sim$100 pc \citep{Frisch2011ARAA}.
This low density cavity in the ISM is thought to be created by a series of supernova explosions in prehistoric time \citep{Zucker2022}. 
The spatial distribution and morphology of the ISM reflects the complex interactions between the gas and stars in the Galactic ``ecosystem", but our knowledge of this neighborhood is far from complete.  For example,  the Loop I bubble is another low density cavity adjacent to the LB, and the well-known giant North Polar Spur (NPS) feature in low-frequency radio sky maps is believed by many to be the brightest part of its bubble shell \citep{Wolleben2007}. However, there are also arguments that the NPS is located at the much more distant Galactic Center and is related to the Fermi Bubble \citep{Sarkar2019}. 

The ISM has a number of different components or phases, which are observed or probed by various methods \citep{Draine2011}.  The atomic and molecular hydrogen, accounting for $\sim80\%$ of the ISM hydrogen in mass,  are traced by the 21cm and CO lines respectively. The dust ($\sim1$\% of ISM in mass) is  measured using interstellar extinction \citep{Lallement2014AA,Capitanio2017AA}. The hot gaseous halo ($\lesssim$ 5\%  of ISM mass within Galactocentric distance of $\sim20$ kpc) is measured
from the X-ray absorption or emission lines of highly ionized metals \citep{Miller2013ApJ}. The  diffuse warm ionized medium (WIM), which occupies $\sim20-30\%$ of the volume near the Galactic plane and accounts for $\sim30\%$ of the total ISM mass \citep{Reynolds1991ApJ, Kulkarni1988gera.book}, has been observed via the diffuse H$\alpha$ or other faint nebular emission lines \citep{Reynolds1977ApJ,Reynolds2000,Haffner2003ApJS,Finkbeiner2003ApJS,Dickinson2003}, or by the pulsar dispersion measures (DM) which could be used to reconstruct the 3D distribution of the electrons \citep{Cordes1991,TaylorCordes1993,Ferriere2001,Gomez2001,Schnitzeler2012MNRAS,Greiner2016AA,Ocker2020ApJ}, though the spatial resolution is limited by the available pulsars. Moreover, the scattering measure (SM) can provide information on the intrinsic  fluctuations  of the WIM \citep{Williamson1972MNRAS}.

The ISM can be probed using low-frequency radio observations, based on the fact that the ionized gas has a stronger absorption of radio waves at lower frequencies. 
The radiative transfer equation of the radiation is 
\begin{equation}
\frac{d I_{\nu}}{ds}= -\alpha_{\nu} I_{\nu} + j_{\nu} 
\end{equation}
where $I_{\nu}$ is the intensity at frequency $\nu$, and the absorption coefficient $\alpha_\nu$ of the ISM at low radio frequency is dominated by the electron free-free process\footnote{The synchrotron self-absorption is unimportant for the typical Galactic magnetic field ($\sim 1~\mu$G) \citep{Sun2008AA} and cosmic ray density ($\sim10^{-13}$ GeV$^{-1}$ cm$^{-3}$) \citep{Peron2021ApJ} above $\sim1$ kHz \citep{Ghisellino2013LNP}. The radio wave cannot propagate below the ISM plasma frequency $\sim2$ kHz \citep{Jester2009NewAR}, so in the frequency range considered in this paper, free-free absorption is the dominant absorption mechanism.}, given by \citep{condon2016essential}
\begin{equation}
\alpha_{\nu} \approx 3.28\times 10^{-7} \left( \frac{T_e}{10^4{\rm K}}\right)^{-1.35} \left( \frac{\nu}{\rm GHz}   \right)^{-2.1} \left(\frac{n_e}{\rm cm^{-3}}\right)^2    {\rm pc}^{-1},
\label{eq:alpha_nu}
\end{equation}
where $T_e$ and $n_e$ are the free electron temperature and number density respectively. For diffuse Galactic electrons the absorption effect becomes significant at $\nu \lesssim10$ MHz, while for dense HII regions the absorption becomes significant at even higher frequencies.
The absorption strength contains information of electron densities at different distances.  By using the multi-frequency data, the density distribution of the electron can be reconstructed, if the radiation emissivity $j_{\nu}$ (or in terms of brightness temperature $T= \frac{c^2 I_\nu}{2 k \nu^2}$, $\epsilon = \frac{c^2 j_{\nu}}{2 k \nu^2}$) is known. 
As this method is sensitive to free electron density, it probes primarily the WIM component of the ISM. Given the large volume and mass fraction of the WIM, it is obviously highly interesting and useful to  reconstruct its 3D structures.

The data at frequencies below 30 MHz, 
which we shall refer to as the {\it ultralong-wavelength} band, and especially below 10 MHz, is very scarce, as ground-based observation is hindered by ionosphere and radio frequency interference (RFI)\citep{Jester2009NewAR}.
There were some ground-based observations at Tasmania of Australia and Canada (e.g. \citealt{Reber1956JGR,Ellis1962Natur,Ellis1966ApJ143,Cane1977MNRAS,Cane1979,Reber1994}, for its history see also  \citealt{low_freq_history_1_2015JAHH,low_freq_history_2_2015JAHH,low_freq_history_3_2015JAHH,low_freq_history_4_2015JAHH,low_freq_history_5_2015JAHH,low_freq_history_6_2016JAHH}); and some early space observation (e.g. \citealt{Hartz1964Natur,Alexander1965ApJ,Smith1965MNRAS,Alexander1969,Brown1973,Alexander1974AJ,Novaco_Brown1978,Manning2001AA}).
From these observations, it is noted that there is a downturn in the global brightness  below $\sim 3$ MHz, which is believed to be caused by the free-free absorption mechanism of the ISM \citep{Ellis1962Natur,Alexander1965ApJ,Ellis1964Natur,Novaco_Brown1978}. 
It has also been noted that at frequencies below a few MHz, in contrast to higher frequencies, the Galactic poles appear to be brighter than the Galactic plane, which is attributed to the stronger absorption on the Galactic plane  \citep{Alexander1970AA,Brown1973,Novaco_Brown1978,Ellis1982AuJPh,Manning2001AA}.

Information on the Galactic electron distribution had been inferred from the ultralong-wavelength observations, even with the crude data obtained in the early observations. 
From the measured ultralong-wavelength global spectrum, by assuming an absorption length, electron density were derived \citep{Hoyle1963AuJPh,Smith1965MNRAS,Ellis1966ApJ146,Alexander1970AA,Brown1973,Cane1979,Kassim1989,Fleishman1995}, which is found to be $\sim0.03-0.1$ cm$^{-3}$ near the Galactic plane. These early works assumed a simple constant electron density. A slightly more sophisticated model assumes that the electron density depends on the vertical distance to the Galactic plane, which is constrained by pulsar DM (e.g. \citealt{Reynolds1991IAUS,Nordgren1992AJ,TaylorCordes1993,Gomez2001}). 
\cite{PetersonWebber2002} found however that in this model the electron density derived from the pulsar observations is too low to give raise the observed downturn of the ultralong-wavelength global spectrum. To fit both observations, they suggested that the electrons in the WIM is clumpy, which would induce stronger free-free absorption.

\cite{Jones2016MNRAS} proposed that the free-free absorption can change the spectral index of the extragalactic background  sources, and this can be used to construct the 2D distribution of electrons (the column density along the line of sight). However, by using discrete point sources as the background, this approach would be limited by the number of available point sources, just as in the pulsar observations.

The existing ultralong-wavelength observations have relatively low resolution. However, reconstruction of the full 3D Galactic electron structures will be feasible when high resolution ultralong-wavelength sky maps are available. Recently, a number of ultralong-wavelength space missions have been proposed \citep{ChenXL2019DSL}, such as the Discovering the Sky at the Longest wavelength (DSL) lunar orbit array \citep{Chen:2020lok,Shi2022a,Shi2022b}, and the Farside Array for Radio Science Investigations of the Dark ages and Exoplanets (FARSIDE) \citep{FARSIDE}, which have the capability of producing high resolution maps, and will enable the full 3D reconstruction.
Anticipating high resolution multi-frequency sky maps in the near future, we propose here to reconstruct the full 3D distribution of electrons using the ultralong-wavelength observation. The spatial distribution of the electrons is encoded in the spectrum:  at the lower frequency, there is a higher contribution of absorption from more nearby absorption. If the emissivity along a line of sight (LoS) is known, then the electron density profile along it can be uniquely determined from a high resolution spectrum, without requiring any prior information about the density profile. Combining the observations of many LoS toward different directions one can obtain the full 3D distribution of the electrons. To our knowledge, such multi-frequency tomographic method has not been applied to the full reconstruction of the ISM before.

The sky radio radiation at different frequencies is produced by various physical processes. Besides the Cosmic Microwave Background  (CMB), the thermal emission of the Galactic dust dominates the radiation above $\sim60$ GHz. The sky between $\sim10-60$ GHz is dominated by the bremsstrahlung radiation from Galactic thermal electrons. In this frequency range, there is also the spinning dust emission that peaks at $\sim20$ GHz. Below $\sim 10$ GHz, the synchrotron radiation produced by the Galactic cosmic ray electrons increases rapidly and becomes dominant \citep{GSM2008MNRAS}.
In the synchrotron background, roughly one-third \citep{Seiffert2011ApJ} is from extragalactic sources such as star-forming galaxies, AGNs, galaxy clusters, and so on \citep{Singal2018PASP}. 
Our proposed reconstruction depends on the observation of the ultralong-wavelength radiation, and the understanding of the unabsorbed synchrotron emissivity. 
The unabsorbed synchrotron radiation has a nearly power-law spectrum, and the variation of the spectral index is relatively small, which simplifies the reconstruction.

The outline of this paper is: In Sec. \ref{sec:methods} we introduce our methods. In Sec. \ref{sec:result} we present the statistics of our results, and
show the reconstructed 1D, 2D (Galactic plane) and 3D structures in simulations and compare them with the input electron model. In Sec. \ref{sec:summary_discussions} we summarize the results and discuss some potential uncertainties. 

\section{Methods}\label{sec:methods}

At the frequency of interest, the emissivity is dominated by the synchrotron radiation from cosmic ray electrons, which are believed to be produced primarily by supernova remnants and other accelerators and propagate over the Galaxy
\citep{Orlando2013MNRAS}. Due to the diffusive propagation, its intensity is relatively smooth, though it can be enhanced near the injecting source or at regions with stronger magnetic field which can trap the particles. Here, we model the emissivity as \citep{Yanping_ULSA}
\begin{equation}
	\epsilon(\nu,R,Z) = A \left(\frac{R+r_1}{R_0}\right)^\alpha
	e^{-R/R_0} e^{-|Z/Z_0|^{\gamma}}\left( \frac{\nu}{\nu_*} \right)^{\beta_{\rm G}}.
	\label{eq:syn_model}
\end{equation}
where $R, Z$ are the Galactic cylindrical coordinates,  
$r_1=0.1$ kpc is a small cut-off radius, $\nu_*=408$ MHz, $A$, $R_0$, $\alpha$, $Z_0$ and $\gamma$  are model parameters, they are obtained by fitting the Haslam 408 MHz map \citep{Haslam1982AA,Remazeilles2015MNRAS}. 
That is, we integrate this emissivity along each LoS to obtain a sky map at 408 MHz, then find the best-fit emissivity parameters for which the sky map is closest (measured by the Euclidean distance) to the observed map. Details can be found in \citet{Yanping_ULSA}. 
We use the constant $\beta_{\rm G}=-2.51$ that is fitted from multi-frequency observations between 45 - 408 MHz.

The purpose of this paper is to demonstrate the feasibility of the algorithm, therefore the simple emissivity model of Eq. (\ref{eq:syn_model}) is used. One may also use more sophisticated models, for example the emissivity constructed by cosmic ray propagation and magnetic distribution  model, and constrained by observations like gamma-ray sky maps, e.g. {\tt GALPROP} \citep{GALPROP_v57}. 
Moreover,  there have been some attempts to construct the synchrotron emissivity from observations. The opaque HII regions can fully absorb the synchrotron radiation behind  them. Therefore they can be used to separate the synchrotron emission from regions in front of and behind the HII regions \citep{Nord2006AJ,Hindson2016PASA,
SuHQ2017MNRAS,SuHQ2018MNRAS,Polderman2019AA}.

\begin{figure*}[t]
	\centering
	 {\includegraphics[width=0.7\textwidth]{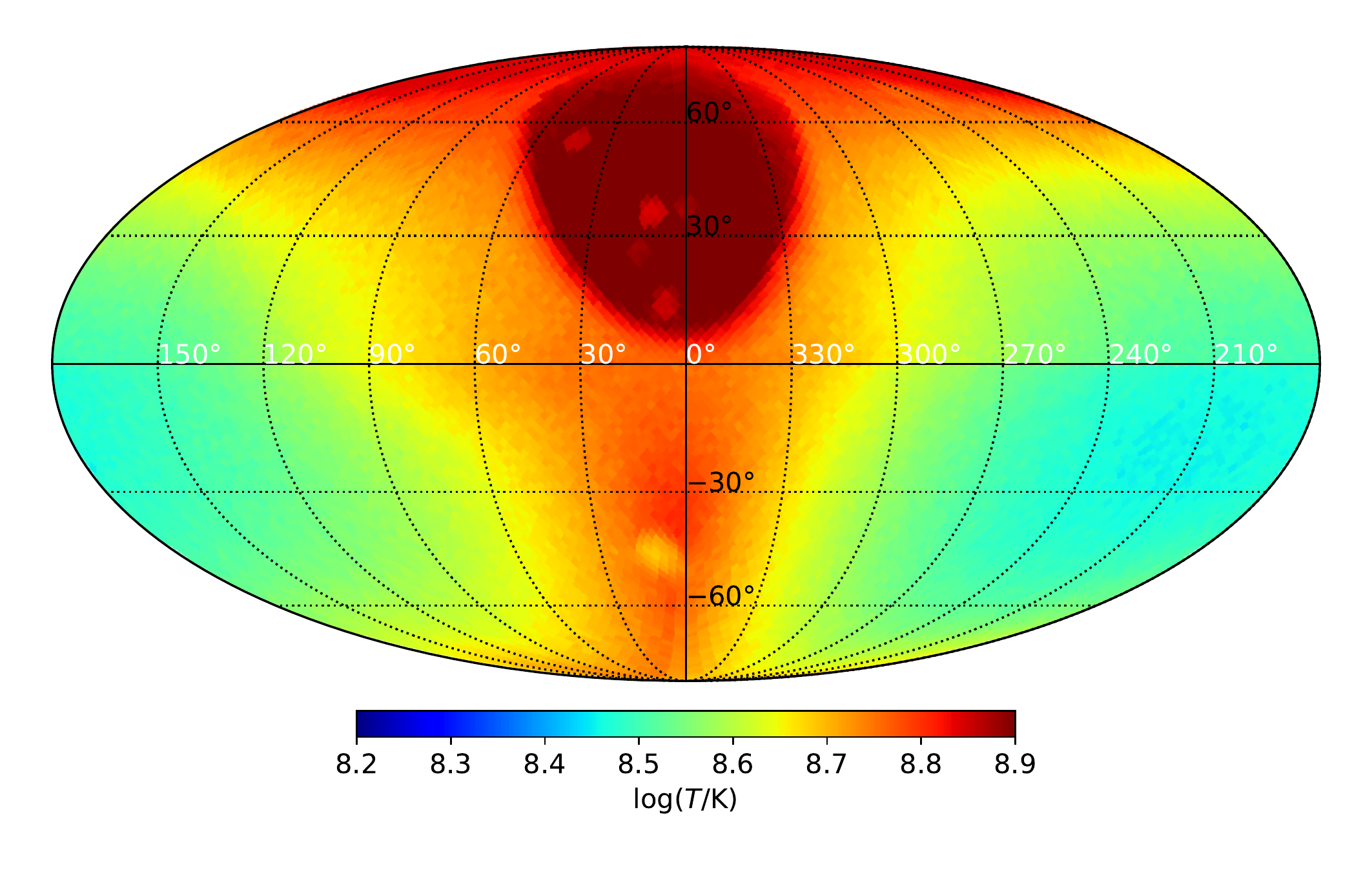}}
	 
	 {\includegraphics[width=0.7\textwidth]{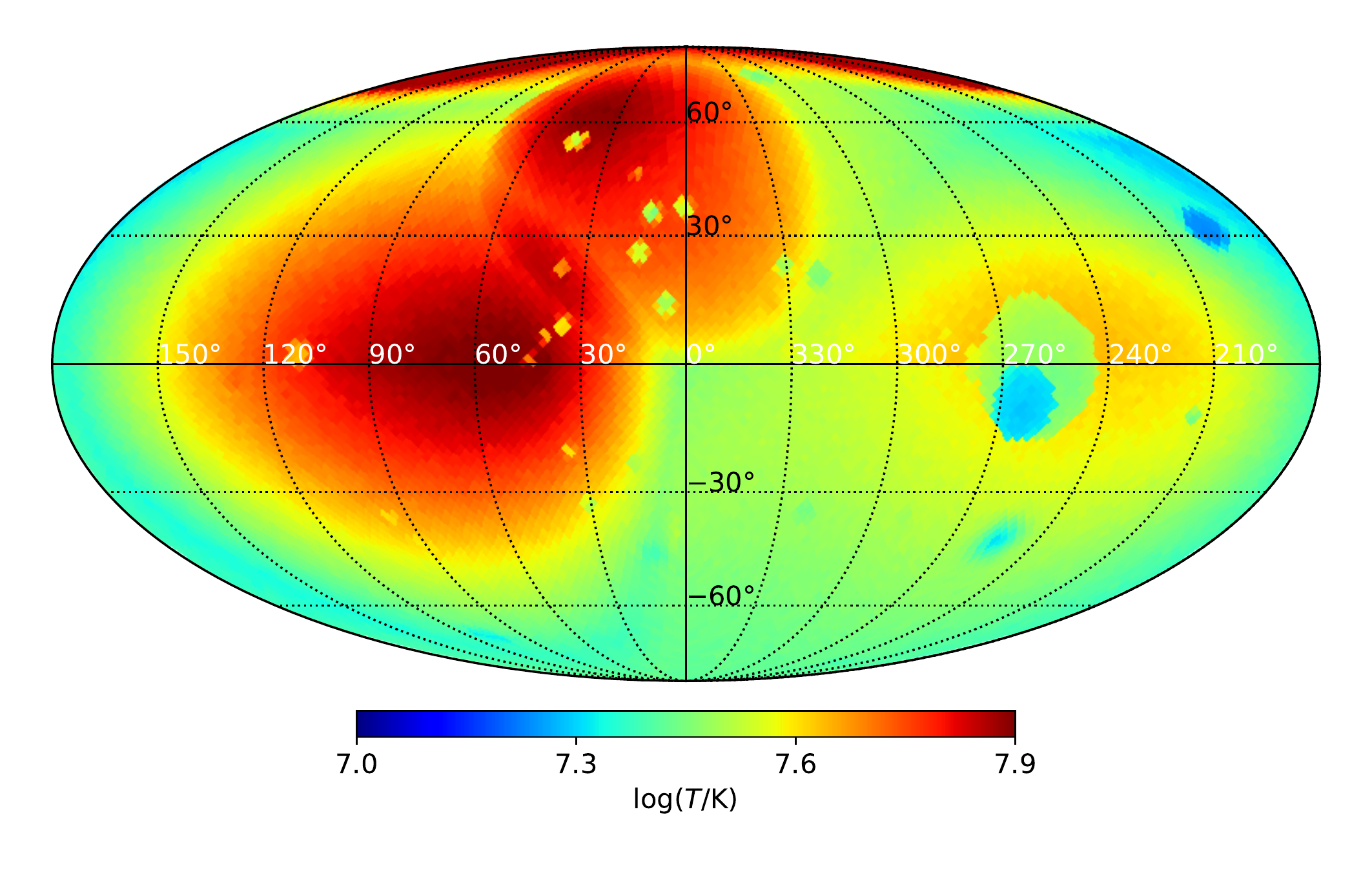}}
	
	\caption{
The mock sky maps at 0.1 MHz (top) and 0.5  MHz (bottom) used in this paper. 
		}\label{fig:skymap}
\end{figure*}

We make the mock multi-frequency sky maps $T_{\rm  sky}^{\rm obs}(\nu_j,l,b)$ using the {\tt ULSA} sky model \citep{Yanping_ULSA}. The {\tt ULSA} generates the sky maps at ultralong-wavelengths including the free-free absorption effect of the Galactic electrons.  There are a number of Galactic electron distribution models by synthesizing different observations 
\citep{Cordes1991,Gaensler2008PASA,
Schnitzeler2012MNRAS,YMW162017}, here the {\tt NE2001} model \citep{ne2001_I,ne2001_II} is adopted. The basic distribution is described by a few components with analytical formulae, for example the thin and thick Galactic disks, and the spiral arms. Known structures such as voids and dense HII regions are also added.
Fluctuations caused by unresolved small-scale ($<1$ pc) electron structures are described by fluctuation parameters. 
When simulating the mock sky maps, we adopt the same \blue{fluctuation parameter} for all the electron components in {\tt NE2001} for simplicity.  
\cite{Yanping_ULSA} found that using the default fluctuation parameters in {\tt NE2001} would over-predict the sky brightness below $\sim 3$ MHz. Instead, if increasing the fluctuation parameter for the thick disk from 0.2 pc$^{-2/3}$ to 3.0 pc$^{-2/3}$, the predicted global spectrum is consistent with the current observations. Since the thick disk dominates the all-sky absorption, we adopt the fluctuation parameter 3.0 pc$^{-2/3}$ for all components.
The produced sky maps  are similar to those in Fig. 8 of \cite{Yanping_ULSA}.
Meanwhile, as our reconstruction relies on the free-free absorption,  adopting a larger fluctuation parameter, 
and hence a more efficient absorption effect, is helpful for reconstructing the electron content. 
Moreover, we adopt a constant electron temperature of 8000 K.
In Fig. \ref{fig:skymap} we plot the mock sky maps $T_{\rm  sky}^{\rm obs}(\nu_j,l,b)$ at $\nu=0.1$ and 0.5 MHz respectively as examples. For sky maps at $\ge 1$ MHz, we refer interested readers to the Fig. 8 of \cite{Yanping_ULSA}.

We demonstrate the reconstruction of electron density distribution from the mock maps. The sky is pixelized  with the {\tt HealPix} scheme, with NSIDE$=$32, so there are a total of 12,288 pixels for the full sky.
Better performance is expected if observations have higher resolutions. 
Along each pixel direction, the LoS is divided into $N$ bins, each with an emissivity and electron density. The emissivity which is dominated by the synchrotron is assumed to be known.  The sky brightness temperature  of this LoS is modeled as 
\begin{align}
T_{\rm sky}(\nu,l,b)\approx &\sum_{i=1}^{N} \bar{\epsilon}_i (\nu,R,Z) \Delta x_i \frac{1-\exp(-\Delta\tau_i)}{\Delta\tau_i} \exp(-\tau_{i-1}) \nonumber \\
&+T^{\rm iso}_E(\nu)\exp(-\tau_N),   
\label{eq:T_lb}
\end{align}
where $\bar{\epsilon}_i$, $\Delta x_i$ and $\Delta\tau_i$ are the mean emissivity, length and optical depth of the $i$-th bin, and 
\begin{equation}
\tau_{i-1}=\sum_{i'=1}^{i-1} \Delta \tau_{i'}.   
\end{equation}
$T^{\rm iso}_E(\nu)$ is the isotropic extragalactic background used in {\tt ULSA} model, given by $T^{\rm iso}_E=1.2(\nu/{\rm GHz})^{-2.58}$ K  \citep{Seiffert2011ApJ}.

We make log-spaced bins along the LoS,  starting from $x_{\rm min}=0.005$ kpc with a step-interval of $\Delta \log x=0.2$, as the brightness temperature is related to the exponential of optical depth. We have limited the LoS to a distance at most 20 kpc from the Galactic center. The
free-free optical depth of each bin is
\begin{align}
\Delta \tau_i &=\alpha_{\nu,i} \Delta x_i \nonumber \\  &= 3.28\times 10^{-7} \left( \frac{T_e}{10^4{\rm K}}\right)^{-1.35} \left( \frac{\nu}{\rm GHz}   \right)^{-2.1} \left(\frac{\mean{n^2_{e,i}}}{\rm cm^{-6}}\right)   \frac{\Delta x_i}{\rm pc}.
\end{align}
If $T_e$ is known, then the reconstructed quantity is $\mean{n^2_{e,i}}$.
Here the symbol ``$\mean{}$'' denotes the mean inside each bin.
From $\mean{n^2_{e,i}}=F_{\rm fluc} \mean{n_{e,i}}^2$, and if the fluctuation parameter\footnote{
When converting the $\mean{n_e^2}$ into $\mean{n_e}$, we define fluctuation parameter $F_{\rm fluc}=\mean{n_e^2}/\mean{n_e}^2$. It is a dimensionless and  phenomenological parameter that describes the connection between the electron density and the free-free absorption strength, without specifying a physical picture of the electron distribution. 
In {\tt NE2001}, the physically motivated fluctuation parameter is defined as $F_{\rm NE2001}=\zeta \omega^2\eta^{-1} l_0^{-2/3}$, where $\omega^2$ (to avoid confusing the variable emissivity, in this paper we replace the ``$\epsilon^2$'' in {\tt NE2001} paper with ``$\omega^2$'') describes the electron fluctuations inside electron clouds, $\zeta$ describes the cloud-to-cloud fluctuations, $\eta$ is the filling factor of electron clouds, and $l_0$ is their outer scale. Therefore $F_{\rm fluc}$ performs as $l_0^{2/3}F_{\rm NE2001} \omega^{-2}(1+\omega^2)$ in formula.
We have checked that when $l_0=1$ pc is adopted as in {\tt NE2001}, the same $F_{\rm fluc}$ and $F_{\rm NE2001}$ values give the same conversion factors. For this reason, we keep the dimensionless definition of the fluctuation parameter as it is only used to display the reconstruction results more intuitively.
} 
$F_{\rm fluc}$ is further known, it is translated into $\mean{n_{e,i}}$. 
For demonstration purpose we assume $T_e=8000$ K \citep{Reynolds1990LNP} and
$F_{\rm fluc}\equiv3.0$ is known, to show the reconstructed electron density as result. Otherwise, our method only reconstructs a combined quantity $T_e^{-1.35}
F_{\rm fluc} \mean{n_{e,i}}^2$. In observations $T_e$ varies between 5500 K and 20000 K \citep{Reynolds1990LNP} or 6000 K - 10000 K (\citealt{Haffner2009RvMP} and references therein), and $F_{\rm fluc}$ can be $\gtrsim 3$ \citep{Gaensler2008PASA}.   
We assume that the emissivity $\epsilon(\nu,R,Z)$ is known and take the form as given 
by Eq. (\ref{eq:syn_model}).

For the multi-frequency maps, we assume the frequency interval is 0.1 MHz for the frequency range of 0.1-1.0 MHz, and 0.2 MHz for the frequency range of 1.0 to 10.0 MHz, so there are totally 55 mock maps at 55 frequencies. We treat $\mean{n_{e,i}}$ of each bin as an unknown parameter. 
By comparing the multi-frequency sky brightness temperature Eq. (\ref{eq:T_lb}) that is a function of $\mean{n_{e,i}}$, 
with observations (mock maps), we can simultaneously
find $\mean{n_{e,i}}$ of all bins that 
minimize
\begin{equation}
\chi^2(l,b)=\sum_{j=1}^{N_{\rm freq}}\frac{ [T_{\rm sky}(\nu_j,l,b)-T_{\rm  sky}^{\rm obs}(\nu_j,l,b)]^2 }{\sigma^2_{\rm noise}},
\label{eq:chi2}
\end{equation}
using the MCMC procedure \citep{EMCEE},
where $N_{\rm freq}$ is the number of frequency bins.

In this work we use fixed grids for solving the electron densities, which may not accurately capture some small and  dense clumps. The absorption by a single dense clump may obscure the whole LoS, but if we already  know some information about these clumps, i.e. their location, size, and density from other observations, we can use such information to improve the fitting.
If a bin contains a  dense clump, then the density contribution from this clump to the bin is known, leaving only the diffuse electron density as an unknown quantity.
For those dense clumps with extremely strong absorption, data at higher frequencies can also be employed.  We assume that clumps with density $>0.1$~cm$^{-3}$ have been  known. For these LoS we add 20 frequency points between 10 and 30 MHz, with an interval of 1 MHz. 
The solution may also be improved with adaptive grids, but we leave such technical improvements to future works.

Regarding the noise level of the observations, at low-frequency the system temperature is dominated by the sky temperature \citep{Shi2022a},
\begin{align}
\sigma_{T_{\rm noise}} &\sim \frac{D^2T_{\rm sky}}{A_{\rm eff}\sqrt{N_a(N_a-1)t_{\rm obs}\Delta\nu}} \nonumber \\
&=\frac{4\pi T_{\rm sky}}{\theta^2_{\rm res}\sqrt{N_a(N_a-1)t_{\rm obs}\Delta\nu}},  
\end{align}
where $A_{\rm eff}=\frac{\lambda^2}{4\pi}$ is the effective area of the antenna with physical size $\ll \lambda$; $N_a$ is the number of antennas; $D$ is the baseline length, and $\theta_{\rm res}\sim\frac{\lambda}{D}$ is the angular resolution; $t_{\rm obs}$ is the integration time and $\Delta \nu$ is the width of the frequency point. 
For $\theta_{\rm res}=1^\circ$, $N_a=8$ and $\Delta \nu=0.1$ MHz, we have $\sigma_{T_{\rm noise}}/T_{\rm sky}\sim 1\%$ if $t_{\rm obs}=0.1$ year; and $\sigma_{T_{\rm noise}}/T_{\rm sky}\sim 10\%$ if $t_{\rm obs}=0.5$ day.
We adopt 1\% noise level for all simulations in this paper.

\section{Results}\label{sec:result}

\begin{figure*}[t]
	\centering
 {\includegraphics[width=0.7\textwidth]{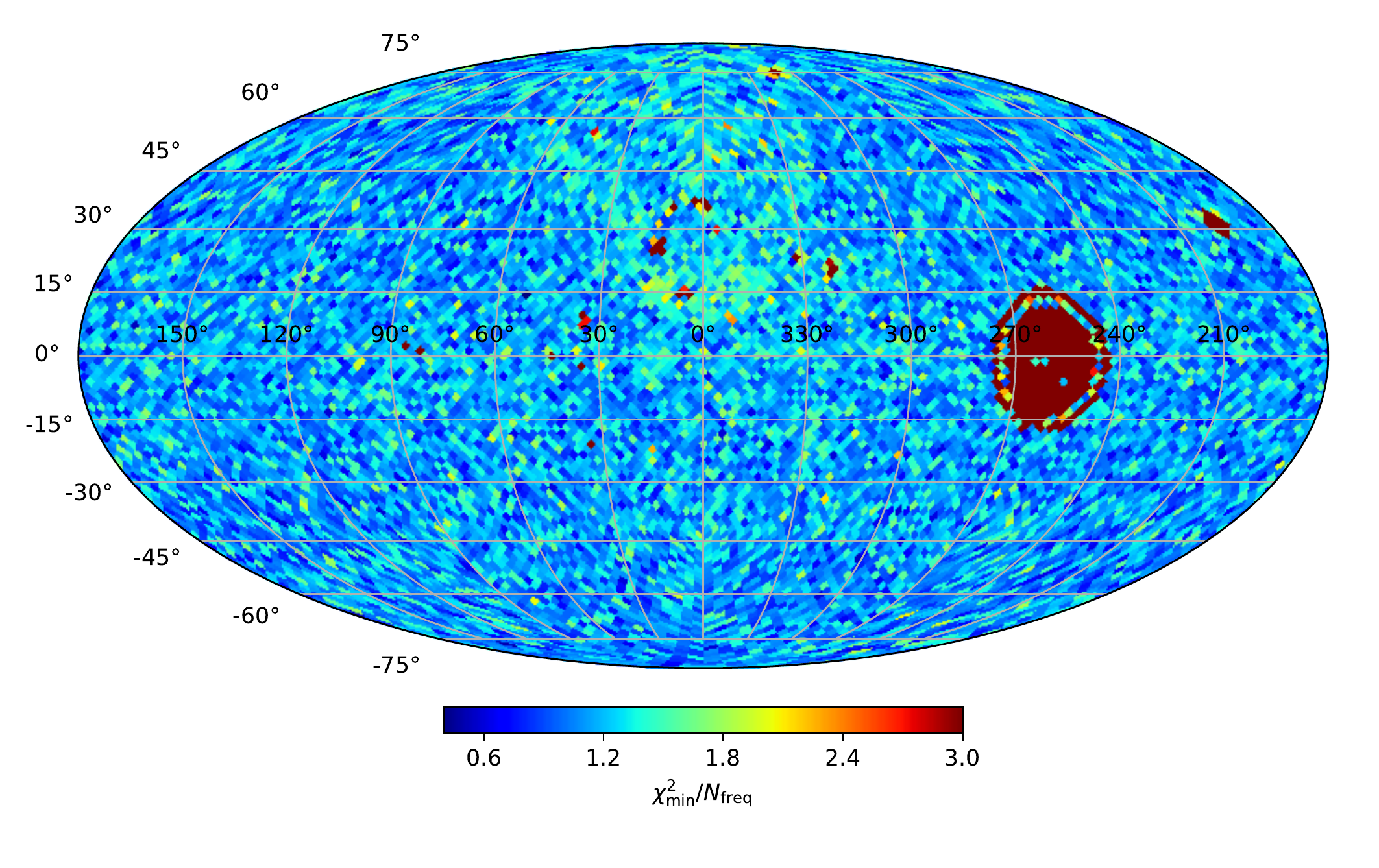}}

{\includegraphics[width=0.7\textwidth]{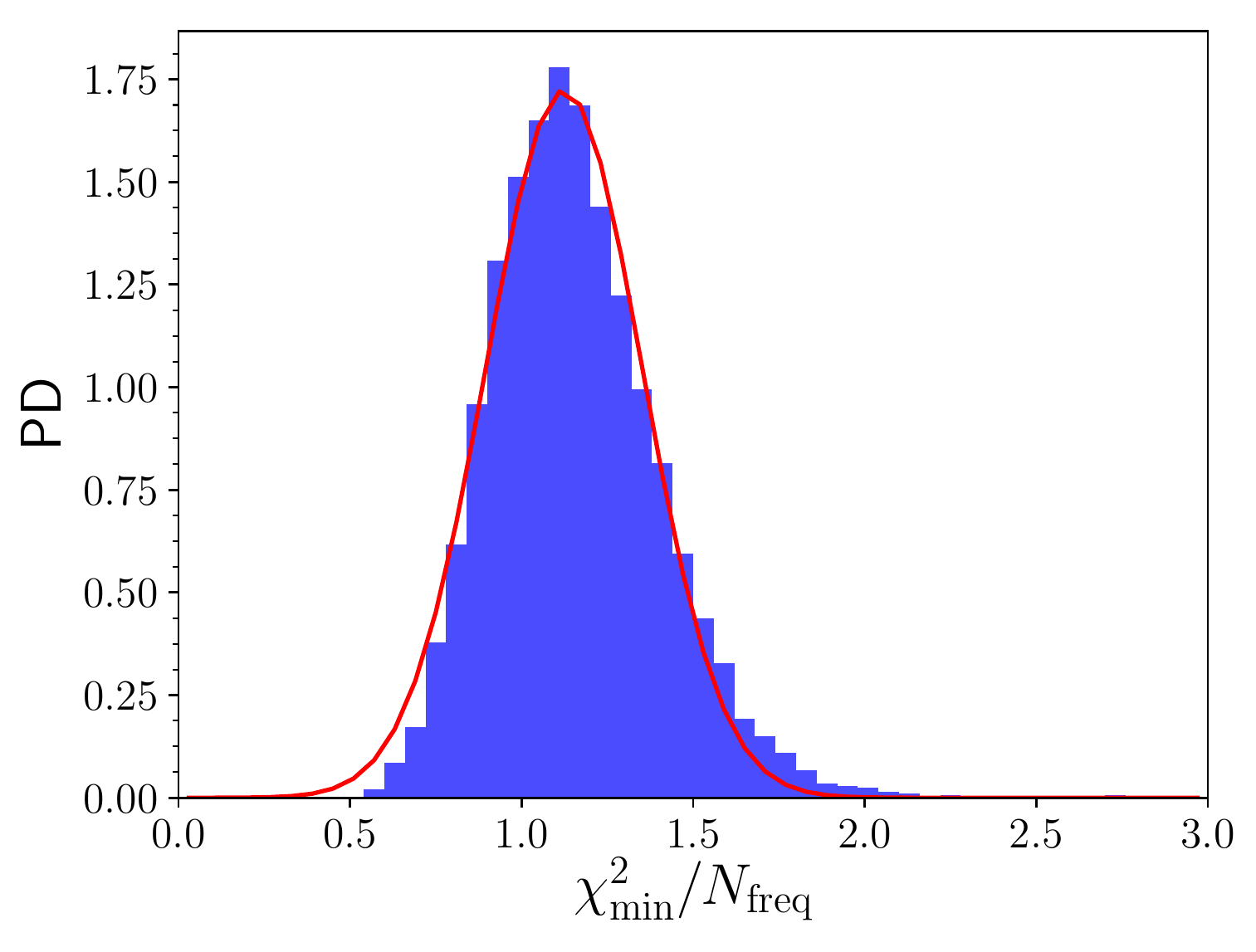}}

	\caption{
	{\it Top:} The sky map of $\chi^2_{\rm min}/N_{\rm freq}$ for all LoS.
 {\it Bottom:} The probability density of the $\chi^2_{\rm min}/N_{\rm freq}$  distribution. 
 The solid curve is a Gaussian fit, with the peak 1.72, the center 1.12 and the radius 0.23.
 There are about $2\%$ of LoS have $\chi^2_{\rm min}/N_{\rm freq}>3.0$.   
 For displaying purpose, in the top panel  they are shown by color same to 3.0, while in the bottom panel these are  not counted.
		}\label{fig:normalized_chi2}
\end{figure*}

In Fig. \ref{fig:normalized_chi2} we show the minimum $\chi^2$ normalized by the number of frequency bins for each LoS as {\tt HealPix} sky map (top), and the probability density (PD) of their distribution (bottom). About 90\% of the LoS have $\chi^2_{\rm min}/N_{\rm  freq}<1.5$, so the MCMC fitting does indeed have a good performance. About 2\% of the LoS have $\chi^2_{\rm min}/N_{\rm  freq}>3.0$, for displaying purpose they are not shown in Fig. \ref{fig:normalized_chi2}. The distribution of $\chi^2_{\rm min}/N_{\rm  freq}$ is well-fitted by a Gaussian function with the peak 1.72, the center 1.12, and the radius 0.23. We check that most LoS with $\chi^2_{\rm min}/N_{\rm  freq}>2.0$ have dense clumps.
As described in Sec. \ref{sec:methods}, if there is a dense clump in a LoS, we 
 assume that its contribution to the density in the relevant bin is known, and only fit the density of diffuse electrons 
 in that bin. This can improve the  reconstructed density profile, see Fig. \ref{fig:1D_LoS} later on. But it cannot solve the problem completely.  Because usually the size of the dense clump is much smaller than the bin width, its absorption to the emissivity in the bin also depends on its exact location inside the bin, but this information is not used in the fix grid computation.   
In the northern hemisphere between $-30^\circ \lesssim l\lesssim 30^\circ$ there is the Loop I bubble, it is a low density cavity with huge angular size, the absorption within the bubble is weak, see the sky maps in Fig. \ref{fig:skymap}. Moreover, such a bubble has an edge that is modeled as a sharp break in the {\tt NE2001}. The sharp break feature is hard to capture because of the bin resolution limit. Because of the above reasons, the $\chi^2_{\rm min}/N_{\rm  freq}$ in this region is biased to higher values.

\begin{figure*}[t]
	\centering
 {\includegraphics[width=0.7\textwidth]{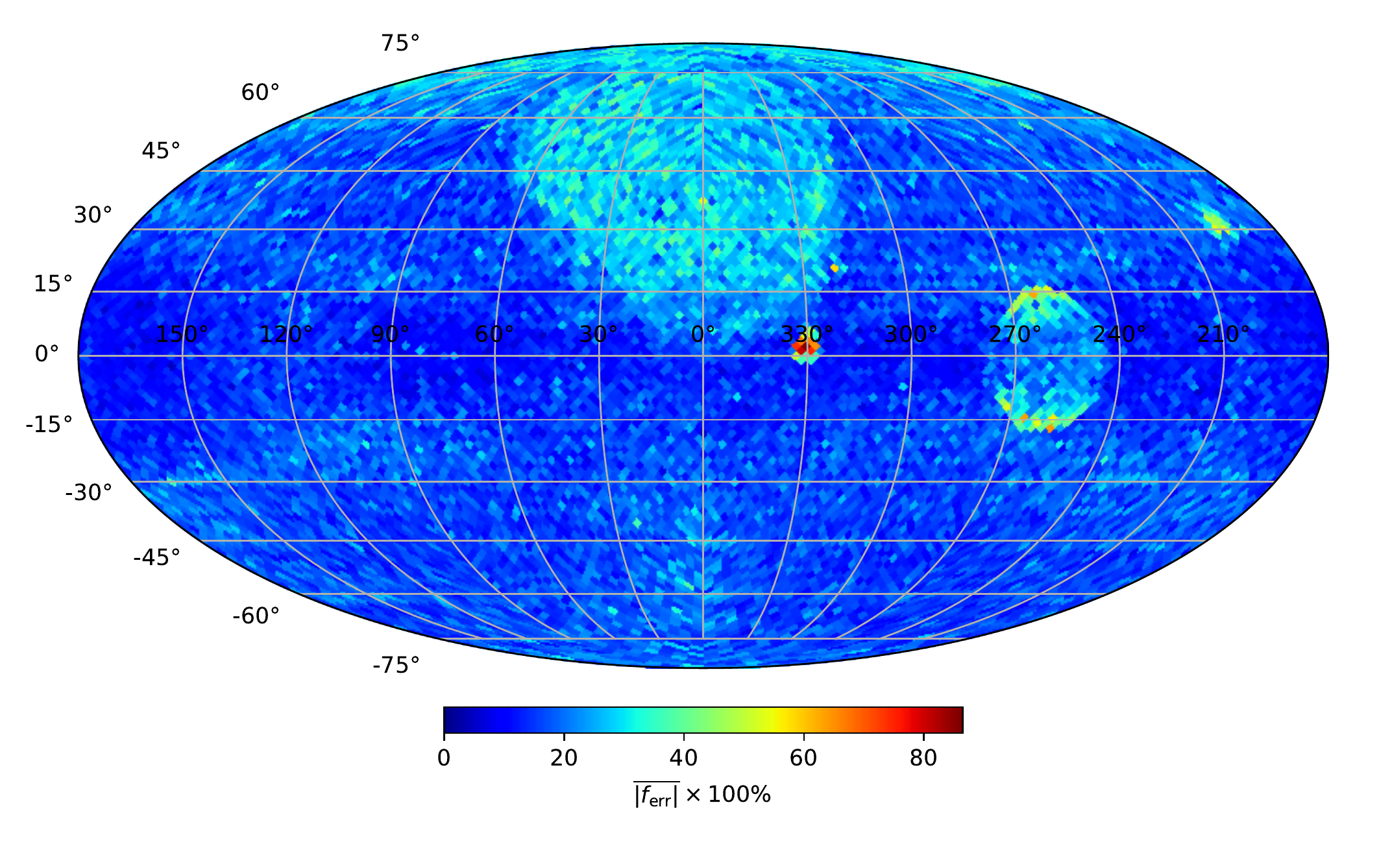}}

{\includegraphics[width=0.7\textwidth]{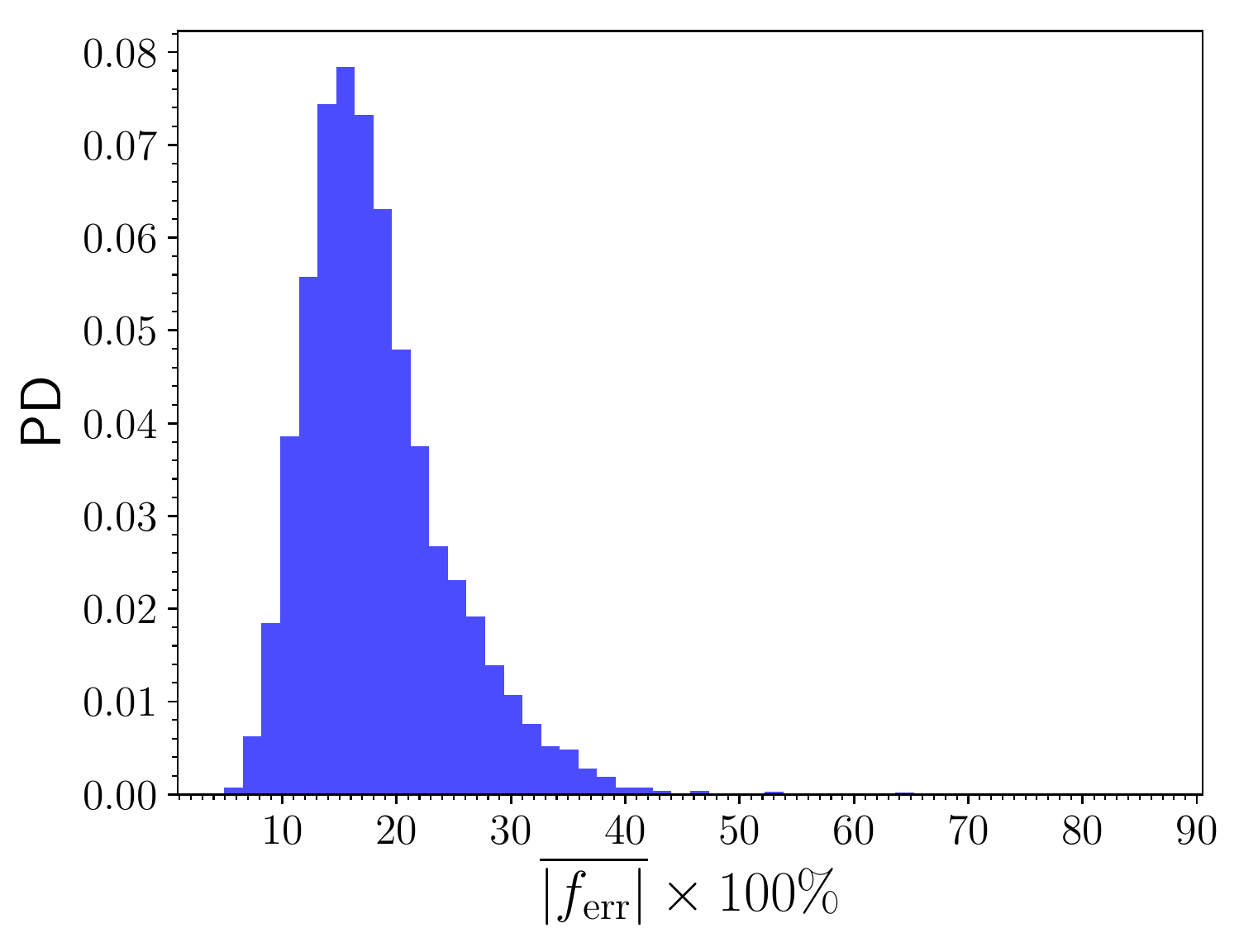}}

	\caption{
	{\it Top:} The sky map for the mean absolute value of the relative error of all LoS. 
 {\it Bottom:} The probability density of the mean absolute value of the relative error distribution. 
		}\label{fig:relative_error}
\end{figure*}

\begin{figure*} 
{\includegraphics[width=0.9\textwidth]{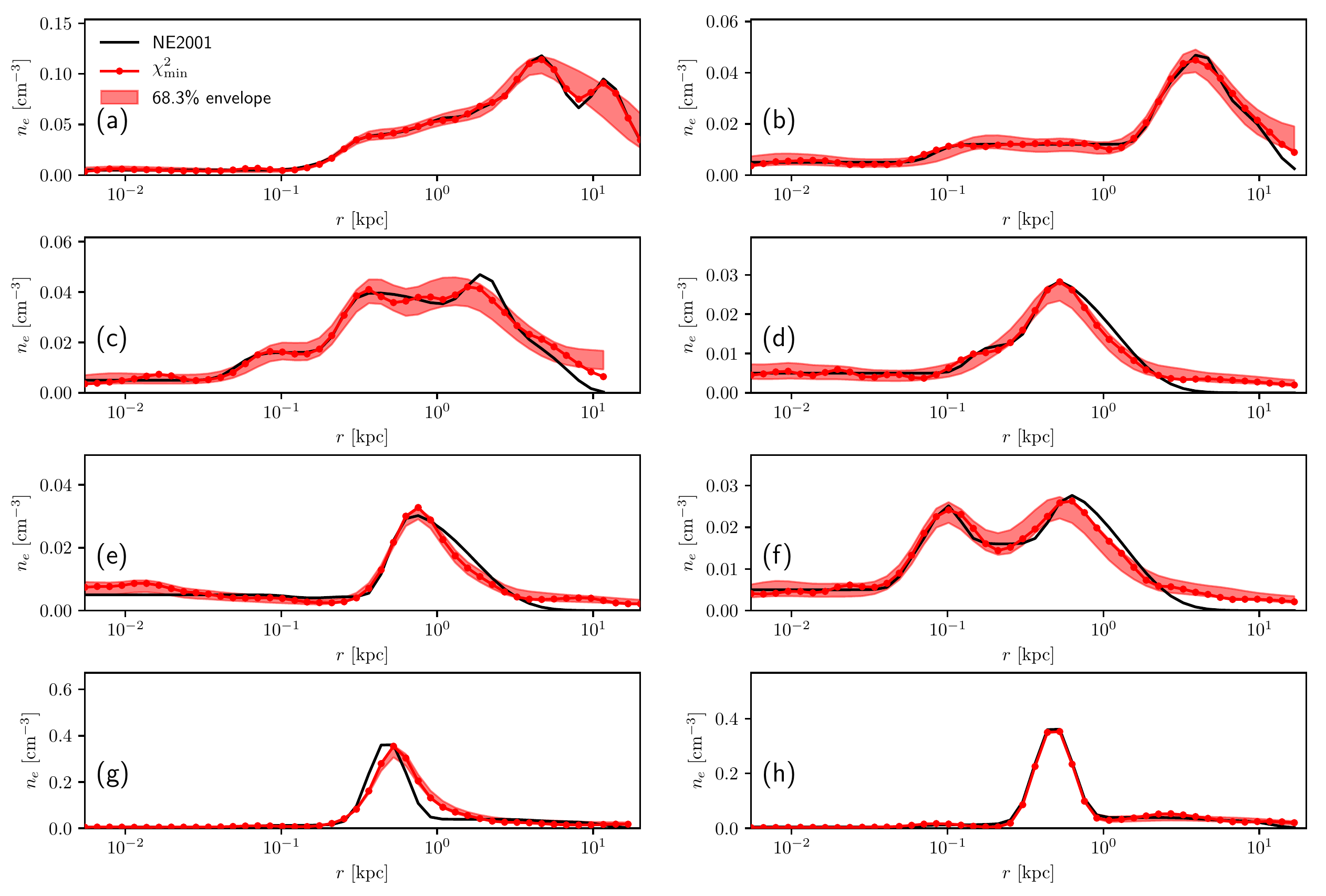}}
\caption{
The reconstructed electron density profiles along 7 LoS in the Galactic plane and in high Galactic regions, pointing to different directions: (a) $l=2^\circ$, $b=0^\circ$; (b) $l=90^\circ$, $b=0^\circ$; (c) $l=180^\circ$, $b=0^\circ$; (d) $l=101.67^\circ$, $b=49.70^\circ$; (e) $l=5.62^\circ$, $b=34.23^\circ$; (f) $l=260.16^\circ$, $b=-41.81^\circ$;
(g) $l=270^\circ$, $b=0^\circ$, the bump at $\sim0.5$ kpc is the Gum Nebula; (h) same as (g) but assume information on Gum Nebula on this LoS is already known, and $0.1 \le \nu \le 30$ MHz data is used for fitting. As comparison in each panel we also plot the  {\tt NE2001}  density  of same $r$ bins. In all panels the reconstructed results and the  {\tt NE2001} model are smoothed by Gaussian kernel with radius equal to one bin.
}\label{fig:1D_LoS}
\end{figure*}

We then compare our reconstructed electron density with {\tt NE2001} 
\footnote{
By definition,  the fluctuation parameter in {\tt NE2001} only accounts for the fluctuations on scales smaller than 1 pc. However, our bin width is much larger than 1 pc. So the fluctuations on scales larger than 1 pc and smaller than bin width are absorbed in the reconstructed quantity  $\mean{n_{e,i}}$ for the $i$-th bin. 
For this bin, if we  divide it into $N_{\rm sub}$ sub-bins with width of 1 pc, then in principle the reconstructed $\mean{n_{e,i}}$ should be more close to
$\sqrt{\sum_j n^2_{e,j}/N_{\rm sub}}$ rather than $\sum_j n_{e,j}/N_{\rm sub}$,
where $n_{e,j}$ is the {\tt NE2001} electron density of the  $j$-th sub-bin in 
the $i$-th bin.
However, in most cases these two quantities are close to each other, unless the bin width $\gtrsim  1$ kpc, or the bin contains  dense clumps.
Throughout this paper, when  comparing with {\tt NE2001} model we take the quantity $\sqrt{\sum_j n^2_{e,j}/N_{\rm sub}}$ of {\tt NE2001}. 
}.
The same fluctuation parameter of 3.0 is assumed for all gas components in order to translate the reconstructed $\mean{n_e^2}$ into the mean electron density that is more intuitive for displaying.
We plot the mean absolute value of relative errors between the reconstructed electron density and that of {\tt NE2001}  
$\overline{|f_{\rm err}|}$ for all LoS as a map in the top panel of Fig. \ref{fig:relative_error}.
To reduce the discrete errors,  both the reconstructed density and the {\tt NE2001} density are smoothed by a Gaussian kernel with a radius equal to one bin width. For each LoS we only count the bins with $R<12$ kpc and $|Z|<2$ kpc, beyond that the electron density is too small and the absorption too weak for relative error to be a good indicator of the reconstruction performance.
In the bottom panel of Fig. \ref{fig:relative_error},  we show the probability density distribution of $\overline{|f_{\rm err}|}$.
About 70\% LoS have $\overline{|f_{\rm err}|}<20\%$. Most of them are near the Galactic plane, see top panel of Fig. \ref{fig:relative_error}. This is natural because in the Galactic plane  the absorption is stronger, making it possible to reconstruct the electron density profile more accurately. High Galactic latitude regions have larger errors, not only because the absorption is weaker, but also because the extragalactic background has a larger fraction in the total received flux.  
Around $l\sim 330^\circ$, there is a patch where the relative error is much higher than the surrounding regions. We check that in {\tt NE2001} there is a low 
 density void with density as low as $0.001$ cm$^{-3}$. In our reconstruction usually the low density regions have larger relative errors. Nevertheless, our reconstruction has good performance, considering the fact that in our method not only each LoS, but also the electron density in each bin of a LoS, are all independent of the others.

We show the reconstructed 1D density profiles for some  LoS in Fig. \ref{fig:1D_LoS} as examples. In Galactic plane our algorithm correctly reproduces the electron density profiles up to $\sim10$ kpc, in high Galactic latitude regions up to $\sim 3$ kpc, although there are deviations on small scales. For all these LoS, within $\sim0.03-0.2$ kpc, the electron density is relatively low, indicating that our Sun is located in a bubble. 
For the LoS pointing to $l\sim 90^\circ$ however, the low density environment extends to $\sim 2$ kpc, this is the 
 low density region simply called ``LDR'' in the {\tt NE2001} model \citep{Toscano1999ApJ,ne2001_II}. 
A LoS penetrating the Gum Nebula is shown in (g) and (h) panels. 
Panel (g) is the result directly reconstructed, and panel (h) is the result when the location, size and density of Gum Nebula is known and its density is added to the relevant bin. Indeed, the density profile in panel (h)  is much improved than in panel (g). 
Nevertheless, except for the Gum Nebula, other dense clumps have much smaller angular sizes on the sky, the large-scale morphology of our reconstructed maps almost does not change, 
if we simply remove those LoS penetrating dense clumps.

\begin{figure*}
\begin{center}
\begin{interactive}{animation}{3D_animation.mp4}
{\includegraphics[width=0.6\textwidth]{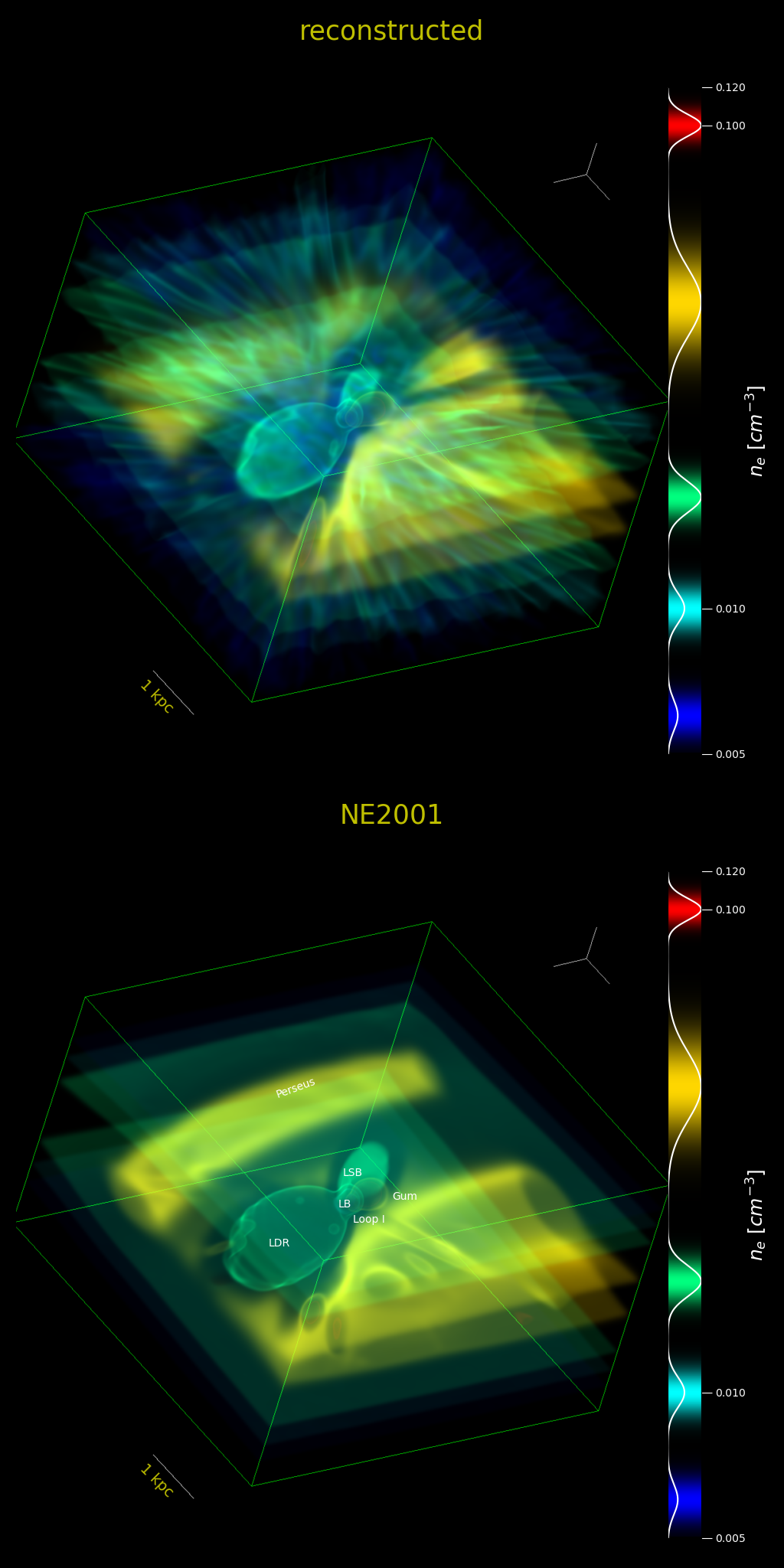}}
\end{interactive}
\caption{
{\it Top:} The reconstructed electron distribution. {\it Bottom :} The {\tt NE2001} model.
Both the top and bottom panel are smoothed by a Gaussian kernel with a radius of 0.05 kpc.
An associated animation for the reconstructed electron density is available online. In the animation, the 3D box rotates about the $Z$-axis in the first 30 seconds duration, so one can see the structures from different perspectives.  In the last 3 seconds, we zoom in the 3D  box until the innermost LB is  seen.  
(On ArXiv, please find the animation in the directory: Ancillary files)
}
\label{fig:3D_visualization}
\end{center}
\end{figure*}

With the mock full-sky map, we can reconstruct the 3D distribution of Galactic electrons by synthesizing all LoS.  The tomographic reconstruction result of our Sun's vicinity is shown in the top panel  of Fig. \ref{fig:3D_visualization}, and for comparison the original {\tt NE2001} model distribution is shown in the bottom panel. 
This figure uses 12288 LoS, and the density volume is a heliocentric cuboid of $6\times 6\times 4$ kpc$^3$, with pixel size $0.05$ kpc.  However, the poorest resolution at the edge is $\sim0.5$ kpc, limited by the bin width.
The reconstructed volume map is also shown in a video linked to Fig. \ref{fig:3D_visualization}.
Both the reconstructed density and the {\tt NE2001} density are smoothed by a Gaussian kernel with radius  0.05 kpc.
Our method reconstructs major features from {\tt NE2001}, including the LB, the LSB, the Gum Nebula, and so on.

In our reconstructed electron density field, there are some streaks as shown in the top panel of Fig.~\ref{fig:3D_visualization}. This is a consequence of our solving the density along the radial direction from our position as the center.  Since we adopt a grid along the radial direction, the grid cells will have shapes that either appear elongated if their size along the radial direction is larger than the tangential direction, or oblate if the reverse is true. In our case we adopted a log-spaced grid which is good for the numerical problem, but at the far end the streak is more apparent. Increasing the number of LoS and reducing the bin width can relieve the steak problem. However, if the bin width is too small then the noise of each bin is large since the contribution of a single bin to the total absorption is small.

\begin{figure*}[t]
	\centering
 {\includegraphics[width=0.8\textwidth]{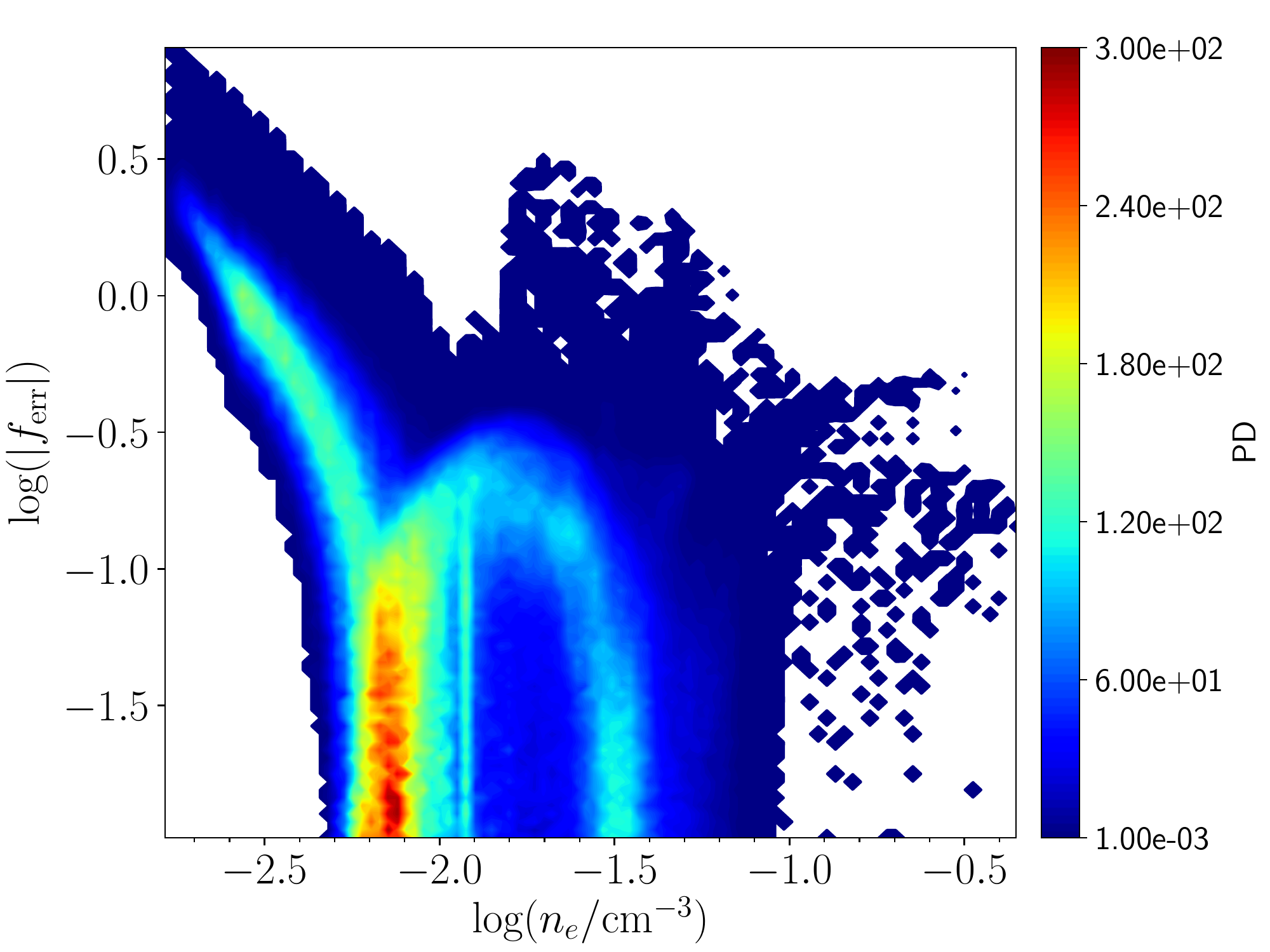}}
	\caption{
The $|f_{\rm err}|$ vs. $n_e$ in the reconstructed 3D Cartesian coordinate density field. Colorbar indicates the probability density with normalization $\int {\rm PD} d |f_{\rm err}| dn_e =1$.   
		}\label{fig:f_err_n_e}
\end{figure*}

In Fig. \ref{fig:f_err_n_e} we show the probability density as a function of $|f_{\rm err}|$ and $n_e$, where $|f_{\rm err}|$ is the absolute value of the relative error between the top  and bottom panels of Fig. \ref{fig:3D_visualization} for each pixel, and $n_e$ is the density of relevant pixel in the bottom panel of Fig. \ref{fig:3D_visualization}.
About 70\% pixels have  $ |f_{\rm err}|< 30\%$, and the median of $|f_{\rm err}|$ is 17\%. About 8\% pixels have $|f_{\rm err}|> 100\%$, most of them are low-density regions, see the top left part of Fig. \ref{fig:f_err_n_e}. This is not surprising because the absorption in low-density regions is weaker, therefore  hard to constrain the density accurately.
Considering the above, any structures with size comparable to the resolution limit and/or with density contrast comparable to the uncertainty level are hard to see from the figure.

\begin{figure*}[t]
	\centering
	 {\includegraphics[width=0.8\textwidth]{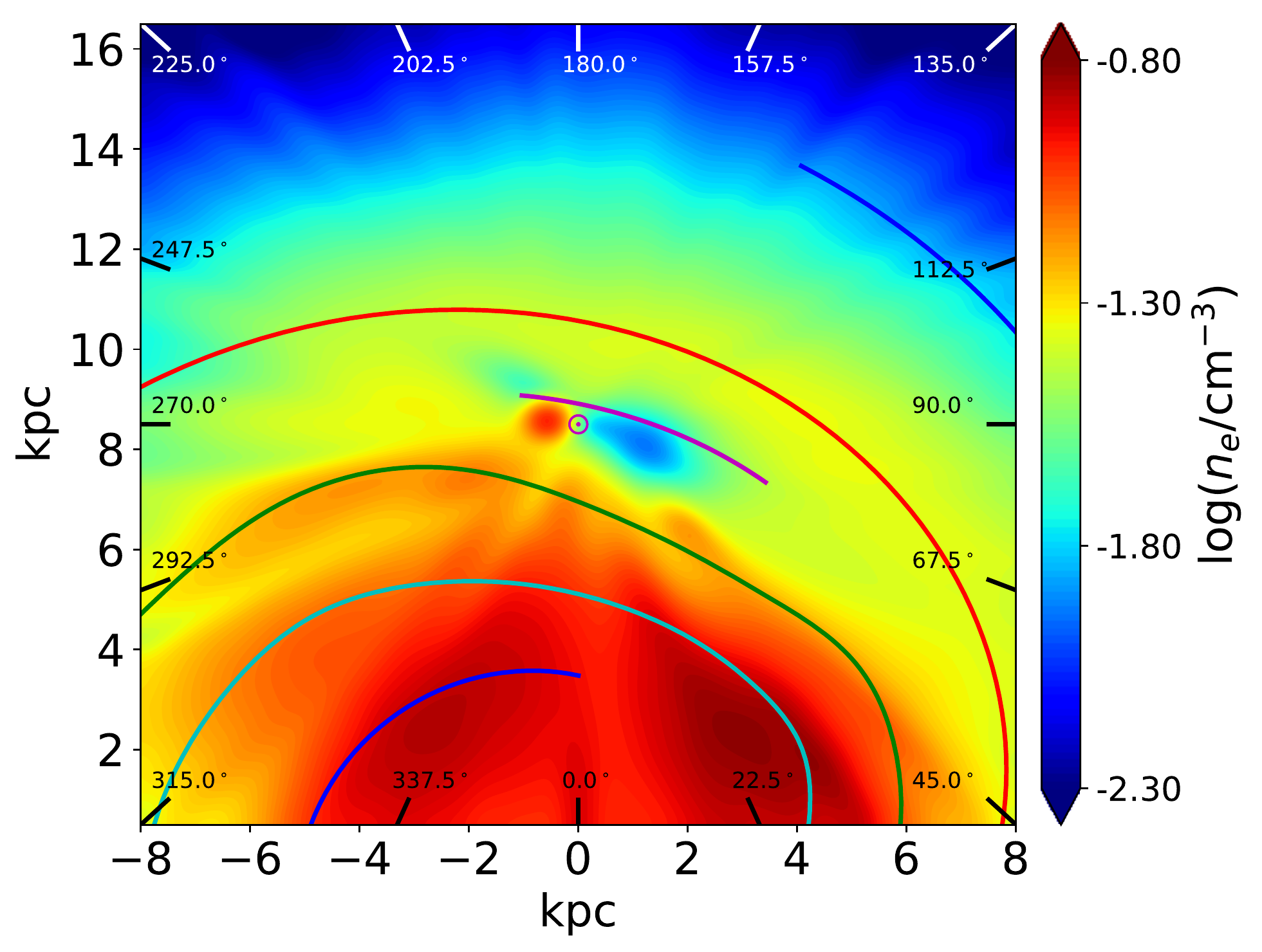}}
	 
	 {\includegraphics[width=0.8\textwidth]{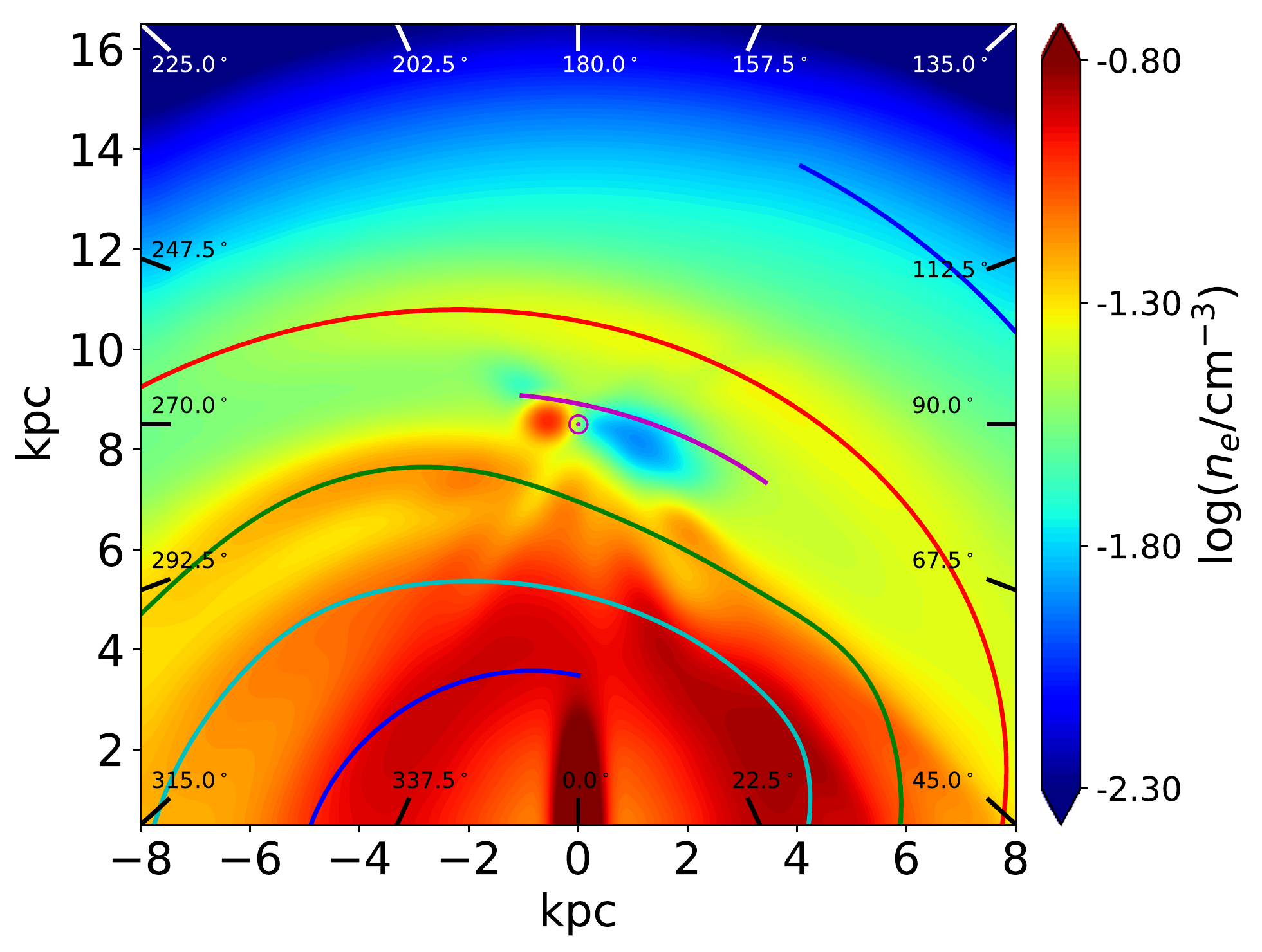}}
	
	\caption{
{\it Top:} The face on view of the electron distribution on  Galactic plane reconstructed from 900 equally separated LoS. To suppress the noise and highlight the large-scale  structures, the density fields are smoothed by Gaussian kernal with a radius 0.25 kpc. {\it Bottom:} The electron density in the NE2001 model. For comparison purpose, the NE2001 data is binned and smoothed same to reconstructed density. 
To guide the eye we plot the locations of spiral arms modeled in {\tt NE2001}. 
		}\label{fig:n_e_Galactic_plane}
\end{figure*}

Moving further from our vicinity on to larger scales, we meet two spiral arms that bracket our Sun, as shown in the reconstructed electron density on the Galactic plane in Fig. \ref{fig:n_e_Galactic_plane}. The reconstructed electron map clearly shows the Carina-Sagittarius arm in the direction towards the Galactic center, and the Perseus arm in the anti-center direction. Previously the spiral arms have been mapped via neutral hydrogen, CO (trace the molecular hydrogen,  \citealt{Englmaier2011MSAIS,Nakanishi2016PASJ}), dust extinction survey \citep{Rezaei2018AA}, and other tracers for high-mass star formation \citep{Hou2014AA}. The Ultralong-wavelength observations will allow  reconstruction of the diffuse free electrons in the spiral arms.  

\section{Summary and Discussions}\label{sec:summary_discussions}

The ultralong-wavelength radio spectrum 
below $\sim$10 MHz has largely been constrained from the ground and space at poor spatial resolution.
Some space missions have been proposed to obtain high-resolution sky maps in this band.
In this paper we proposed that, thanks to the frequency-dependent absorption to electromagnetic wave, from the upcoming ultralong-wavelength observations the three-dimensional structures of the Galactic electrons can be reconstructed. Using mock sky maps we performed simulations to prove the feasibility of this method.  

There are dense clumps in the sky, and
they often have active star formation and also  emit synchrotron radiation. If such emission is not accurately known, then it may cause errors in reconstruction. To assess this uncertainty, we consider the nearby Gum Nebula. According to \cite{Woermann1998PhDT}, at  408 MHz the Gum Nebula has typical surface brightness temperature of roughly 10 K, and the typical spectrum index derived from 408 and 2326 MHz is roughly $-2.5$.  Taking the Gum Nebula as a sphere with radius 0.14 kpc and electron density 0.43 cm$^{-3}$, at a distance 0.5 kpc from us \citep{ne2001_I}, and the emissivity is uniformly distributed in the sphere, we expect $\sim 5\times10^4$ K surface brightness at 1 MHz if $F_{\rm fluc}=3.0$. This is actually smaller than the mean brightness of the Galactic plane in our sky model at the same frequency. 
According to this simple estimate, a large fraction of the clump emission might be absorbed by itself, since it has density much higher than the ambient medium. So at least in this case, the influence of the clump emission is modest.

\begin{figure*}
\centering
 \includegraphics[width=0.9\textwidth]{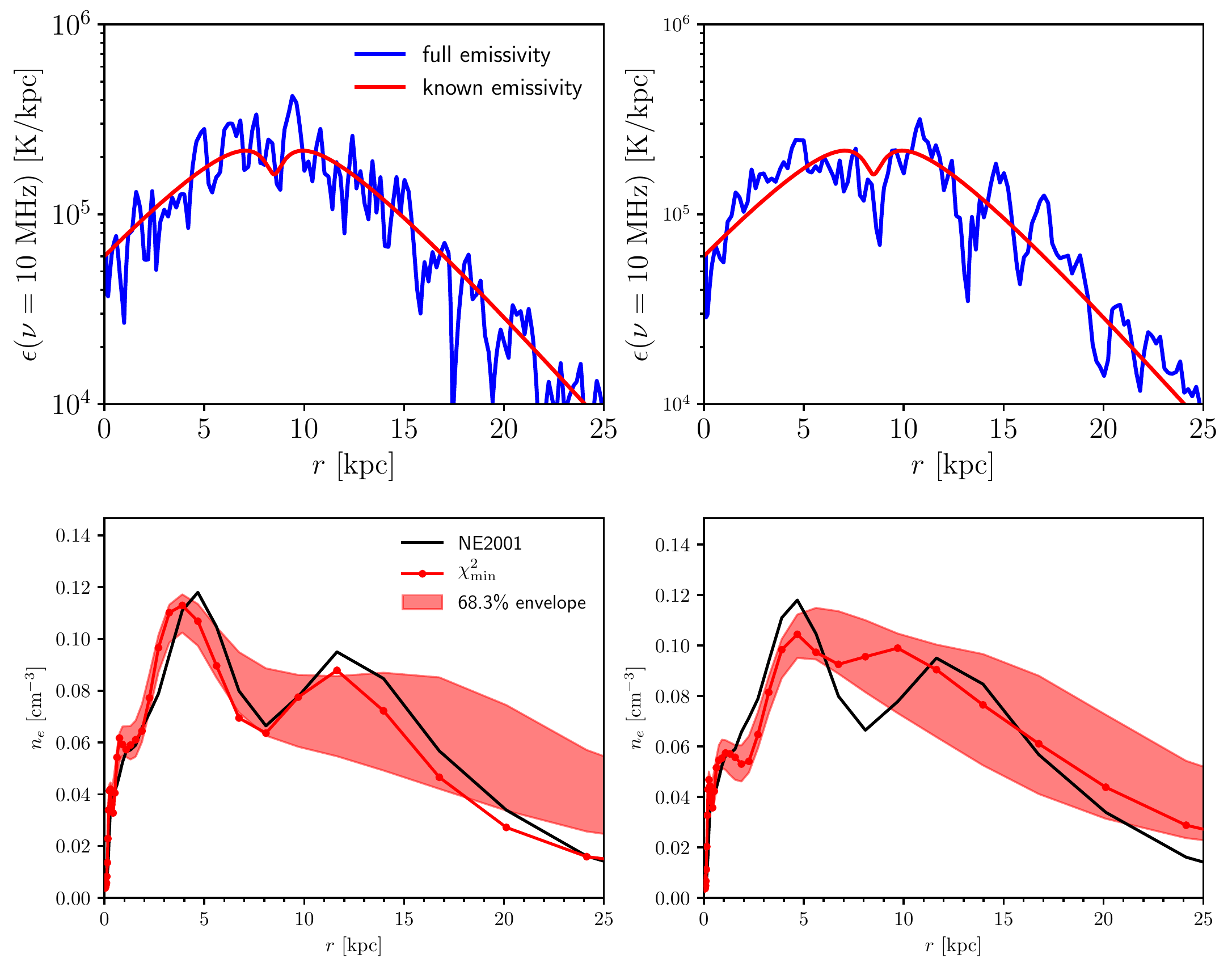}
\caption{ 
The emissivity with small-scale fluctuations and its known smooth component (top panels), and the reconstructed electron density profiles for the LoS with $l=2.0^\circ$ on Galactic plane (bottom panels).  
For left panels we adopt $k_{\rm min}=5.0 $ kpc$^{-1}$  and for right panels we adopt $k_{\rm min}=0.5$    kpc$^{-1}$. All density profiles, including the NE2001, have been smoothed by a Gaussian kernel with radius equal to one bin.
}\label{fig:emiss_fluc}
\end{figure*}

Our synchrotron emissivity model is cylindrical. However the real emissivity could have non-cylindrical large-scale structures, induced by the spiral structures in the regular magnetic field \citep{Orlando2013ICRC}. We have tested that for such a spiral emissivity model, the electron density can be still reconstructed and all  conclusions do not change. The reconstruction errors could come from the unknown random small-scale fluctuations in the synchrotron emissivity. 
To estimate its impact, we add fluctuations to the emissivity model of Eq.~(\ref{eq:syn_model}), and then reconstruct the electron density profile assuming the small-scale fluctuations are not present. Motivated by  the measured angular power spectrum of Galactic synchrotron radiation  $C_l\propto l^{-2}$ 
\citep{Iacobelli2013AA},  we model the small-scale fluctuations of emissivity to be Gaussian with the power spectrum 
\begin{equation}
P_\epsilon(k) \propto k^{-3}, ~~ k  > k_{\rm min}, 
\label{eq:P_k}
\end{equation}
where $k$ is the spatial wavenumber. We assume that fluctuations only exist on scales smaller  than $2\pi/k_{\rm min}$ ($k> k_{\rm min}$). 
On larger scales, we assume the Eq.~(\ref{eq:syn_model}) correctly describes the cosmic ray emissivity.
We normalize
$P_\epsilon$ so that the emissivity has a mean relative fluctuations level $\sim50\%$ at the smoothing scale 
0.1 kpc.
That is, when the fluctuations  field following the power spectrum of Eq. (\ref{eq:P_k}) is added to the cylindrical
emissivity field, and the new emissivity field is smoothed with Gaussian kernel with radius $=0.1$ kpc, then $\sqrt{\overline{(\delta \epsilon)^2}
}/\overline{\epsilon}\sim 50\%$.
We fond that, despite this perturbation, which induces errors in the reconstruction at small scales, 
the reconstruction result remain largely unchanged at large 
scales. 
The results of this reconstruction are shown in Fig. \ref{fig:emiss_fluc}, for $k_{\rm min} = 5.0~\mathrm{kpc}^{-1} $ and $k_{\rm min} = 0.5 ~\mathrm{kpc}^{-1} $ respectively.

Our method directly reconstructs $\mean{n_e^2}$ of each bin, a quantity that produces absorption equivalent to real electron distribution inside this bin. 
Since $\mean{n_e^2}=F_{\rm fluc} \mean{n_e}^2$, we actually cannot distinguish the cases if a region has smaller electron density but larger fluctuation parameter, or the reverse, unless the $F_{\rm fluc}$ is derived from other methods. Just for convenience of translating the results into a more intuitive quantity $\mean{n_e}$ we assume a constant $F_{\rm fluc}$. The $F_{\rm fluc}$ may vary bin-by-bin since the fluctuations of the WIM  can depend on locations \citep{Reynolds1991ApJ,Gaensler2008PASA}, however we have checked that it will not change our reconstruction of $\mean{n^2_e}$. Just, if the correct $F_{\rm fluc}$ is not known, the translated $\mean{n_e}$ would deviate from the real mean electron density.  
For example, in \cite{Gaensler2008PASA} they investigated the volume filling factor of electron clouds. If the electron fluctuations inside clouds are negligible, then the fluctuation parameter can be well approximated as the reciprocal of the filling factor.
In this case the fluctuation parameter
is $\sim25$ in the Galactic plane and $\sim3$ at $|Z|=1.4$ kpc. For such a fluctuation parameter then our $\mean{n_e}$ at Galactic plane is overestimated by a factor $\sim3$. 
In \cite{Ocker2021ApJ} they considered a complicated electron cloud model rather than uniform density. According to the observations of scattering and dispersion measures of the Galactic pulsars, they found that the fluctuation parameters at the low Galactic latitudes and inner Galaxy are $\gtrsim 10-1000$ times larger than the typical thick disk at high latitudes.
Considering this, the error on the reconstructed electron density by assuming a constant fluctuation parameter can be even much larger, particularly at low Galactic latitudes and for  the LoS towards the inner Galaxy.
Future studies trying to disentangle the electron density would merit a more sophisticated model for the spatial variation in the fluctuation parameter.
Perhaps values independently inferred from pulsar observations along similar LoS could be employed, although this would be limited by the sparseness of pulsar spatial distribution. 
In some biased regions, the WIM has larger fluctuations. The synchrotron emissivity from such regions will be also biased, because usually such regions are more close to the sources of the cosmic ray\---  supernova remnants. 
We tested that, if the biased  emissivity is correctly involved in the emissivity model, then no matter if the fluctuation parameter is biased or not in the same  regions, the $\mean{n^2_e}$ can still be correctly reconstructed.
However if the biased emissivity is missed in the model, then like what we have discussed in last paragraph, it would  induce small-scale uncertainties on the reconstructed $\mean{n^2_e}$ and the final $\mean{n_e}$.   

Despite some modeling uncertainties, the reconstruction from ultralong-wavelength observation is fairly robust.
Once the ultralong-wavelength sky is systematically surveyed, we will gain  new knowledge about our residence in the Milky Way, and pin down the long-lasting debate on the distance of the NPS. It will be an important step in
obtaining
a synthetic picture of our Galaxy. 

\section*{Acknowledgments}
We thank the anonymous referee for the very helpful comments.
 This work is supported by National SKA Program of China Nos. 2020SKA0110402 and 2020SKA0110401, Inter-government cooperation Flagship program grant No. 2018YFE0120800, the National Natural Science Foundation of China grant Nos. 11973047, 11633004, the Chinese Academy of Sciences (CAS) Key Instrument grant ZDKYYQ20200008, the Strategic Priority Research Program XDA15020200, and the CAS Frontier Science Key Project QYZDJ-SSW-SLH017.
 
\bibliography{mybib}{}
\bibliographystyle{aasjournal}



\end{document}